\documentclass[runningheads]{llncs}

\usepackage{amssymb}
\usepackage{amsfonts}
\usepackage{graphicx}
\usepackage{subcaption}
\usepackage{hyperref}
\usepackage{xspace}
\usepackage{algorithm,algorithmic}

\usepackage[utf8]{inputenc}
\usepackage{xcolor}

\newcommand{\ring}[1]{\ensuremath{\mathbb{#1}}}
\newcommand\KK{\ring{K}}
\newcommand\QQ{\ring{Q}}
\newcommand{\gb}{Gr\"obner basis\xspace}
\newcommand{\gbs}{Gr\"obner bases\xspace}

\begin{document}
\title{Analysis of the Conradi-Kahle Algorithm for Detecting Binomiality
  on Biological Models \thanks{This work has been supported by the
    bilateral project ANR-17-CE40-0036/DFG-391322026 SYMBIONT. We
    would like to thank the authors of \cite{luders_odebase:_2019} for
    providing us with the polynomials that we used in our computations.}}
\titlerunning{Analysing the Conradi-Kahle Algorithm}
\author{Alexandru Iosif\inst{1} \and Hamid Rahkooy\inst{2}}
\institute{JRC-COMBINE Aachen, Germany \and CNRS, Inria and the
  Universit\'e de Lorraine, Nancy, France \\
  \email{iosif@aices.rwth-aachen.de}\\
  \email{hamid.rahkooy@inria.fr}}
\maketitle 
\begin{abstract} We analyze the Conradi-Kahle Algorithm for detecting
  binomiality.  We present experiments using two implementations of
  the algorithm in Macaulay2 and Maple on biological models and assess
  the performance of the algorithm on these models. We compare the two
  implementations with each other and with \gbs computations up to
  their performance on these biological models.

\end{abstract}

%===================Intro=====================
\section{Introduction}  

We study the problem of binomiality of polynomial ideals. Given an
ideal with a finite set of generators, we would like to know if there
exists a basis for the ideal such that its elements have at most two
monomials. Such an ideal is called a \textit{binomial ideal}.  We use
the Conradi-Kahle Algorithm for testing whether an ideal is
binomial. Our investigations are focused on implementing this
algorithm and performing computations.  Binomial ideals offer clear
computational advantages over arbitrary ideals. They appear in various
applications, e.g., in biological and chemical models.

Binomial ideals have been extensively studied in the literature
\cite{eisenbud1996binomial,kahle2012decompositions,kahle2014positive}.
Eisenbud and Sturmfels in \cite{eisenbud1996binomial} have shown that
\gbs \cite{Buchberger:65a} can be used to test binomiality. Recently,
biochemical networks whose \textit{steady state ideals} are binomial
have been studied in the field of \textit{Algebraic Systems Biology}
\cite{craciun_toric_2009,gatermann_bernsteins_2005,millan2012chemical}.
Mill\'an and Dickenstein in \cite{millan_structure_2018} have defined
\textit{MESSI Biological Systems} as a general framework for
modifications of type enzyme-substrate or swap with intermediates,
which includes interesting binomial systems
\cite{millan_structure_2018}.

In the context of biochemical reaction networks, Mill\'an,
Dickenstein, Shiu and Conradi in \cite{millan2012chemical} present a
sufficient condition on the \textit{stoichiometric matrix} for
binomiality of the steady state ideal. Conradi and Kahle
\cite{conradi2015detecting} proved that this condition is necessary
for homogeneous ideals and proposed an algorithm. The Conradi-Kahle
Algorithm is implemented in Macaulay2 \cite{BinomialsPackage}. Iosif,
Conradi and Kahle in \cite{Conradi2019} use the fact that the
irreducible components of the varieties of binomial ideals admit
monomial parametrization in order to reduce the dimension of detecting
total concentrations that lead to multiple steady states.

Our contribution in this article is analysing efficiency and
effectiveness of the Conradi-Kahle Algorithm, using Gr\"obner bases
for reduction, applied to some biological models. We first discuss the
complexity of the algorithm and reduce it to the complexity of
computing a \gb for a preprocessed input set of polynomials. Then we
present our computations in Macaulay2 \cite{Macaulay2} and Maple
\cite{Maple} and compare the algorithm with simply computing \gb of
the input ideal which shows the strength of the algorithm. The
experiments are performed on biological models in the BioModels
repository \footnote{\url{https://www.ebi.ac.uk/biomodels/}}, which is
a repository of mechanistic models of bio-medical systems
\cite{BioModels2015a,BioModels2018a}. Our intial motivation was to
understand the advantages and disadvantages of the method in
\cite{millan2012chemical} for testing binomiality of chemical reaction
networks. As the Conradi-Kahle Algorithm follows the idea of the
method in \cite{millan2012chemical} with more subtle reduction steps,
we rather use the Conradi-Kahle Algorithm to check binomiality of
ideals coming from biomodels, although none of our steady state ideals
are homogeneous.

%================The Conradi-Kahle Algorithm==============
\section{The Conradi-Kahle Algorithm}\label{sec:maple}

The Conradi-Kahle Algorithm is based on the sufficient condition by
Mill\'an, Dickenstein, Shiu and Conradi \cite{millan2012chemical} for
binomiality of steady state ideals. The latter states that if the
kernel of the stoichiometric matrix has a basis with a particular
property then the steady state ideal is binomial. Conradi and Kahle
converted this into a sufficient condition for an arbitrary homogenous
ideal $I$ generated by a set $F$ of polynomials of fixed degree. They
proved that $I$ is binomial if and only if the reduced row echelon
form of the coefficient matrix of $F$ has at most two non-zero
elements in each row. This leads to the Algorithm \ref{alg:CK} which
is incremental on the degrees of the generators.

\begin{algorithm}
  \caption{(Conradi and Kahle, 2015)}
  \label{alg:CK}
  \begin{algorithmic}[1]
  \REQUIRE Homogeneous polynomials
  $f_1,\ldots,f_s\in\KK[X]$, where $\KK$ is a field.\\
  \ENSURE \emph{Yes} if the ideal $\langle f_1 , \ldots, f_s \rangle$
  is binomial. \emph{No} otherwise.
    \STATE Let $B:=\emptyset, R:=\KK[x_1,\ldots,x_n]$ and
    $F:=\{f_1,\ldots,f_s\}$.
    \WHILE {$F \ne \emptyset$}
    \STATE Let $F_{\min}$ be the set of elements of minimal degree in
    $F$.
      \STATE  $F := F \setminus F_{\min}$.
      \STATE  Compute the reduced row echelon form $A$ of the
      coefficient matrix of $F_{\min}$. 
      \IF{ $A$ has a row with three or more non-zero entries}
       \RETURN \emph{No} and stop
      \ENDIF
      \STATE Let $M$ be the vector of monomials in $F_{\min}$.
      \STATE Let $B'$ be the set of entries of $AM$.
      \STATE  $B := B \cup B'$.
      \STATE  $R := \KK[x_1, \ldots,x_n]/\langle B \rangle$.
      \STATE  Redefine $F$ as the image of $F$ in $R$.
      \ENDWHILE
    \RETURN \emph{Yes}.
  \end{algorithmic}
\end{algorithm}

Now we analyze the complexity of Algorithm~\ref{alg:CK}.  
\begin{itemize}
  \item \textbf{Steps $3$ and $4$.} can be ignored. 
  \item \textbf{Step $5$.} Let $t$ denote the number of distinct
    monomials in
    $F_{\min}$ and $m:=\max(s,t)$. Computing the reduced row echelon
    form of $A$ can be done in at most $m^{\omega}$ steps, where
    $\omega$ is the constant in the complexity of matrix multiplication. 
  \item \textbf{Step $6$.} needs at most $st$ operations which is less
    or equal than $m^{\omega}$, so we ignore this term.
  \item \textbf{Steps $10$.} can be bounded by $tm$, which itself can
    be bounded by $m^{\omega}$, hence ignored.
   \item \textbf{Step $12$.} This can be done  via computing a \gb of
     $\langle B \rangle$. Another way to do this, is by means of
     Gaussian elimination on the corresponding Macaulay matrix of $B$.
  \item \textbf{Step $13$.} is equivalent to reducing $F$ modulo
    $\langle B \rangle$, which can be done via reducing $F$ modulo a
    Gr\"obner basis of $\langle B \rangle$. Another method to do this
    is via Gaussian elimination over the Macaulay matrix of $F \cup B$. 
\end{itemize}

Following Mayr and Meyer's work on the complexity of computing \gbs
\cite{mayr_complexity_1982}, computations in steps $11$ and $12$ of
the algorithm can be EXP-SPACE. Conradi and Kahle observe through
experiments that these steps can be performed via graph enumeration
algorithms like breadth first search, which makes it more efficient
than Gr\"obner bases in practice \cite{conradi2015detecting}. In this
article we do not use such graph enumeration algorithms in our
implementations. This is the subject of a future work.

%==============M2 & Maple Experiments================
\section{Macaulay2 and Maple experiments}

We consider 20 Biomodels from the BioModels repository
\cite{BioModels2015a,BioModels2018a} whose steady state ideal is
generated by polynomials in $\QQ(k_1,\dots,k_r)[x_1,\dots,x_n]$ where
$k_1$, $\dots$, $k_r$ are the parameters and $x_1,\dots,x_n$ are the
variables corresponding to the species. Our polynomials are taken from
\cite{luders_odebase:_2019}. We use Algorithm~\ref{alg:CK} to test
binomiality of these biomodels. We emphasise that in our computations
we do not assign values to the parameters $k_1,\dots,k_r$ and we work
in $\QQ(k_1,\dots,k_r)[x_1,\dots,x_n]$. We have implemented Algorithm
\ref{alg:CK} in Maple \cite{MapleCK} and also use a slight variant of
the implementation of the algorithm in the Macaulay2 package Binomials
\cite{BinomialsPackage,kahle2012decompositions}. We also test
binomiality of an ideal given by a set of generating polynomials via
computing a \gb of the ideal, using Corollary 1.2 in
\cite{eisenbud1996binomial}. Our computations are done on a 3.5 GHz
Intel Core i7 with 16 GB RAM. In our computations we used Macaulay2
1.12 and Maple 2019.1.

  \begin{table}[htpb]\centering
\scalebox{1.0}{
    \begin{tabular}{ | c || c | c | c | c |c|c|c|}
      \hline
      Biomodel & C-K (M2) & C-K (Maple) & Bin (C-K) & GB (M2) &
                                                    GB (Maple) & Bin (GB) \\
      \hline
      2 & 0.1 & 1 & No && &   \\ \hline
      9 &0.04&0.2 & Yes & 0.5 &0.001 &Yes\\\hline
      28 &0.04&0.1 & No&  & & \\\hline
      30 &0.5&0.2 &No& & & \\\hline
      46 &0.02&0.2 &No&100 &80 &No\\\hline
      85 &0.04&0.6 & No& & & \\\hline
      86 &0.08&6 &No&  & & \\\hline
      102 &0.04&0.2 &No&  & &\\\hline
      103 &0.1&0.9 &No& & & \\\hline
      108 &0.01&0.03 &No&  & & \\\hline
      152 &0.3&400&No& & & \\\hline
      153 &0.4&500& No& & & \\\hline
      187 &0.02&0.07&No&0.06 & 0.1 &No \\\hline
      200 &0.05&1&No& & & \\\hline
      205 &0.6&50& No& & & \\\hline
      243 &0.04&0.3&No& 0.01 & 0.05 &No \\\hline
      262 &0.05&0.02&Yes&0.01 & 0.02& Yes\\\hline
      264 &0.7&0.03& Yes&2& 0.04& Yes\\\hline
      315 &0.02&0.2& No&& & \\\hline
      335 &0.04&0.8&No&30&90 &No \\\hline    
    \end{tabular}
  }
  \caption{
    \label{fig:computations}
    CPU times (in seconds) for Algorithm \ref{alg:CK} and \gbs.}
\end{table}

Table~\ref{fig:computations} shows the results of our computations.
Biomodel columns in the table shows the number of the biomodel. The
columns C-K (M2) and C-K (Maple) show the CPU timings in seconds of
executing Algorithm \ref{alg:CK} in Macaulay2 and Maple,
respectively. In the column Bin (C-K), Yes means that the algorithm
successfully determined that the ideal is binomial, while No means
that the algorithm cannot determine whether the ideal is binomial or
not. The columns GB (M2) and GB (Maple) are the timings of \gbs
computations of the input polynomials in Macaulay2 and Maple,
respectively. The Macaulay2 and Maple timings are rounded to the first
nonzero digit. Bin (GB) column is blank if the \gb computation did not
finish after 600 seconds. Yes in the latter column means that \gb
computation finished and shows that the ideal is binomial, while No
shows that the \gb computation finished but it detected that the ideal
is not binomial.

None of the ideals in the biomodels that we have studied are
homogeneous. Therefore, in order to use Algorithm~\ref{alg:CK} we need
to homogenise the ideals. Consequently, if the algorithm returns
\emph{No}, we are not able to say whether the ideal is binomial or not
(see~\cite[Section~4]{conradi2015detecting}). As one can see from the
column Bin (C-K), the Conradi-Kahle Algorithm is able to test
binomiality only for Biomodels $9$, $262$ and $264$. If \gbs
computations finish, then they can test binomiality for every
ideal. However, as one can see from the related columns, this is not
the case. Actually in most of the cases, \gbs computations did not
finish within 600 seconds. One can see from the table that whenever
\gbs computations give a yes answer to the binomiality question, then
the Conradi-Kahle Algorithm also can detect this as well. In the Yes
cases, the timings for both methods in both Macaulay2 and Maple are
very close.

Algorithm~\ref{alg:CK} returns the output within at most a few
seconds, however, most of the Gr\"obner bases computations did not
finish in 600 seconds. The advantage of testing binomiality using \gbs
computations can be seen in Biomodels $46$, $187$, $243$ and $335$,
where \gbs computations---although slower---show that the ideal is not
binomial, but the Conradi-Kahle Algorithm cannot detect this in spite
of its fast execution. With a few exceptions, we do not observe
significant difference between Macaulay2 and Maple computations,
neither for the Conradi-Kahle Algorithm nor for the \gbs
computations. We would like to emphasise that the Conradi-Kahle
Algorithm is complete over homogeneous ideals. However, in this
article we are interested in ideals coming from some biological models
which are inhomogeneous, and this might affect the performance of the
algorithm. In future we will do experiments on homogeneous ideals in
order to better understand the performance of the algorithm in that
case.\\

\noindent\textbf {Acknowledgement.} We would like to thank the anonymous
referees for their comments.

%
% ---- Bibliography ----
%
%\bibliographystyle{splncs04}
%\bibliography{binomiality-ck-gb}

%=======BBL FILE========

%========
\end{document}